%% file: ms.tex
\pdfoutput=1
\documentclass[12pt, a4paper, oneside]{article}
\usepackage[utf8]{inputenc}
\usepackage[english]{babel}
\usepackage{textcomp}
\usepackage{cmap}
\usepackage[top=2cm, bottom=6cm, left=2cm, right=2cm, includefoot, heightrounded]{geometry}  % <- keep this space!
\usepackage{nicefrac}
\usepackage{amsmath}
\usepackage{hyperref}
\usepackage{morefloats}
\usepackage{placeins}

\usepackage{helvet}
\usepackage{times}
\usepackage{booktabs}
\usepackage{comment}
\usepackage{fancyhdr}
\usepackage{titlesec}

\usepackage{graphicx}
\usepackage[table,xcdraw]{xcolor}
\usepackage[alf, abnt-emphasize=bf]{abntex2cite}
\usepackage{subfig}
\usepackage[T1]{fontenc}
\emergencystretch=\maxdimen
\titlespacing*{\section}{0pt}{0.25\baselineskip}{0.25\baselineskip}
\titlespacing*{\subsection}{0pt}{0.25\baselineskip}{0.25\baselineskip}
\titlespacing*{\subsubsection}{0pt}{0.25\baselineskip}{0.25\baselineskip}

\titleformat*{\section}{\large\bfseries} 
\titleformat*{\subsection}{\large}
\titleformat*{\subsubsection}{\normalsize}
\begin{document}
% --------------------------------------------------------------------------
% Title of the paper
% --------------------------------------------------------------------------
	\begin{center}
		\large
		\textbf{Practical Analysis of Permeable Concrete Properties with Polypropylene Fiber Addition} \\
		\vspace{0.2cm} % <- keep this space!
		\normalsize
		\vspace{0.5cm} % <- keep this space!
	\end{center}
% --------------------------------------------------------------------------
% Authors
% --------------------------------------------------------------------------
\begin{center}
	Rebeca de M. Kich (1); Victor A. Kich (2); Kelvin I. Seibt (3)
	
	\vspace{0.5cm}  % <-keep this space!
	
    \footnotesize 
	\textit{(1) Universidade do Vale do Taquari - UNIVATES} \\
	\textit{Student, B.Sc., Civil Engineering}\\
    \textit{rebecademkich@gmail.com}\\
    \textit{(2) Universidade Federal de Santa Maria - UFSM} \\
	\textit{Student, B.Sc., Computer Engineering}\\
    \textit{victorkich@yahoo.com.br}\\
    \textit{(3) Universidade de Santa Cruz do Sul - UNISC} \\
	\textit{Student, B.Sc., Computer Engineering}\\
    \textit{kelvin.seibt@gmail.com}
\end{center}
\vspace{0.5cm}  % <- keep this space!

% -----------------------------------------------------------------------
% Abstract
% ----------------------------------------------------------------------

\noindent {\large\textbf{Abstract}} \par
{\small
\noindent 
\input{sections/0_abstract}
}
\noindent{\fontfamily{ptm} \selectfont \textit{keyword: Concrete mixes, Permeable concrete, Polypropylene fiber, Mechanical properties.}}

\clearpage
\input{sections/1_introduction}
\input{sections/2_methodology}
\input{sections/3_results}
\input{sections/4_conclusion}

\newpage
\renewcommand\ref{name{\vspace{-1.5cm}}}
\bibliographystyle{abntex2-alf}
\bibliography{referencias}

\end{document}

%% file: sections/0_abstract.tex
%\begin{abstract}

One of the recent approaches used to minimize the impacts of the growth of impermeable areas in urban centers is permeable flooring. Permeable floors can be made of concrete and are called permeable concrete. The objective of this research is to analyze the use of polypropylene fibers in the mixture of permeable concrete to assess whether the fibers significantly alter the properties of this type of concrete, such as compression resistance and flexure tensile strength. For the tests, three concrete mixes of permeable concrete were first used to determine the reference of the composition of concrete: 1:3, without the use of sand and with water-to-cement ratio (w/c) equal to 0.32; 1:4, with the use of 10\% of coarse sand and w/c of 0.35; and finally 1:5, with the use of 10\% of coarse sand and w/c of 0.35; and finally 1:5, using 10\% of coarse sand and w/c of 0.35. In these concrete mixes, tests were performed to determine their mechanical properties being them: permeability, compression resistance, and flexure tensile strength. After the determination of the reference concrete mix, three more concrete mixes were made with the use of polypropylene fibers, at levels of 0.6 kg/m$^3$, 1.8 kg/m$^3$, and 3.0 kg/m$^3$. The same initial tests were then carried out for concrete with the addition of fibers. The analysis of the results showed that the concrete mix that achieved the best result for the flexure tensile strength test, which was the main focus of the research, was the 1:4 trait with the addition of 1.8 kg/m$^3$.

%\end{abstract}

%% file: sections/1_introduction.tex
\section{Introduction}

With the acceleration growth of metropolises in the last decades, the extent of permeable ground territory has decreased, changing their natural drainage system. Aiming to the decrease of problems caused by these changes, are being adopted the Best Management Practices (BMPs) methods, which have the purpose of avoiding the transference of upstream problems to downstream hydrographic basins.

\citeonline{virgiliis2009procedimentos} found that one of the principal devices utilized for urban drainage is the permeable pavement. It has the capacity to absorb the rainwater, transporting it to a gravel reservoir below it.

The permeable concrete is characterized by having a high index of empty spaces interconnected, having in its constitution a large number of coarse aggregates and small or none quantity of fine aggregates, allowing the drainage of large volumes of water. However, due to its porosity, mechanical resistance has relatively low values if compared with the conventional concretes \cite{fagundesavaliaccao}. 

According to \citeonline{abreu2015concrete}, the addition of fibers to concrete is directly related to its durability, as concrete has a low flexure tensile strength when compared to its compression resistance. In this way, it ends up generating cracks when the concrete part is tested. The fibers help to prevent these cracks from occurring. Evaluating this context, it can be said that almost no fibers are used in permeable concrete because the main problem of this type of concrete is not cracking, but its low resistances.

Polypropylene fibers have a large capacity of deformation, huge resistance to alkalinity, and a low cost. Though, those fibers have a low resistance to fire, are sensitive to solar light, and a limited grip on the concrete matrix. They are recommended when is wanted to control the cracks that occur during concrete hardening, which makes them recommended to use in locals that have huge surfaces, such as pavements \cite{oliveira2016analise}.

According to \citeonline{fonseca2016concrete}, the utilization of polypropylene fibers in the permeable concrete helps in the reduction of cracks and increase the mechanical resistance so that the concrete supports greater efforts. The large capacity of deformation that this type of fiber has improves in the decrease of cracks.

\begin{figure}[ht]
\centering
\includegraphics[width=0.7\textwidth]{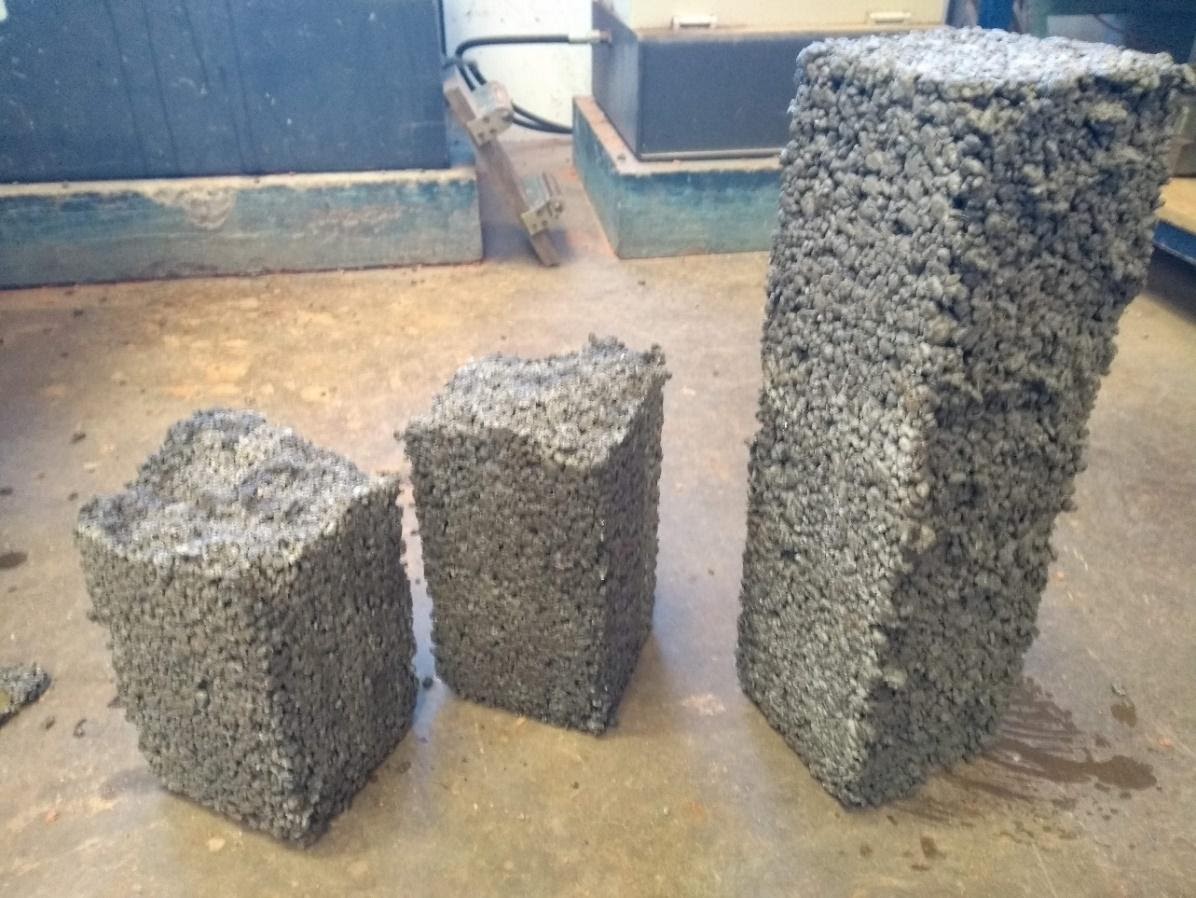}
\caption{Specimens where the resistance tests were performed.}
\label{CORPOS}
\end{figure}

The general objective of this research is to analyze the influence of the addition of polypropylene fibers (in different quantities) in the mechanical performance and physical properties of permeable concrete, doing practical tests in specimens properly built. These specimens are shown in figure \ref{CORPOS}. Thus, the results of two distinct environments were compared: one where had permeable concrete mix without the addition of any fiber, and another with permeable concrete mix added of polypropylene fibers.

%% file: sections/2_methodology.tex
\section{Methodology}

To analyze the influence of polypropylene fiber addition, two environments with tree permeable concrete mixes were considered, where the first testings were without the utilization of fiber. The characterized material for the experiments was: cement CP - ARI, coarse sand, gravel nº 0, kneading water, and polypropylene fiber. The present study will be considered the execution of permeable concrete slabs to light traffic.

The cement used in the tests was the cement \textit{Portland} of high early strength CP V - ARI with a specific mass of 2.72 g/cm$^3$. The choice of this cement has been followed by the determinations of NBR 16697 \cite{associaccao2018nbr}.

The fine aggregate that has been used was the thick quartz sand from natural origin. To test the specific mass were is used the procedures as NBR NM 52 \cite{nbr2009agregado}. This test consists of weighing a sample of 500 g of the aggregate in a dry state and put it in a container, where is added water, staying the sample submerged for 24 hours. After that, it is withdrawal and scattered on a surface, where must stay over the effects of an air current to dry naturally. Then, undergoes the sample to 25 blows in a specific container, who is withdrawn, and this should cave in. If the sample does not cave in, the test is repeated until this happens.

To test the unitary mass were is used the procedures as NBR NM 45 \cite{nbr2006agregado}. This parameter is found by dividing the mass of the aggregate by the aggregates volume. The results found to the specific mass and unitary mass of the fine aggregates are shown in Table \ref{aggregates}.

\begin{table}[ht]
\centering
\caption{Aggregates.}
\label{aggregates}
\begin{tabular}{l c c} 
\toprule
\textbf{Type} & \textbf{Specific mass (g/cm$^3$)} & \textbf{Unitary mass (g/cm$^3$)} \\
\midrule
Fine aggregate & 2.67 & 1.59\\
Coarse aggregate & 2.49 & 1.50\\
\bottomrule
\end{tabular}
\end{table}

The coarse aggregate that was used in this research was the coarse nº 0, following the \citeonline{fagundesavaliaccao} and \citeonline{monteiro2010concreto}, having used 9.5 mm for maximum aggregate sizes.

To test the specific mass of the coarse aggregate were is used the procedures as NBR NM 53 \cite{nbrnm532009concrete}. The test consists of weighing a sample of coarse aggregate and put it in a container added of water, staying the sample submerged for 24 hours. Then, the sample is wrapped in a cloth during the time necessary to dry the surface of the aggregates. A new weighing is done, and after a weighing with the sample submerged in the water. The unitary mass was determined in the same way that in the fine aggregate, as NBR NM 45 \cite{nbr2006agregado}. The results found to the specific mass and unitary mass of the coarse aggregates are shown in Table \ref{aggregates}.

The water used was from the public supply of the region. As NBR 15900 \cite{nbr1590002009concrete}, the water from public supply does not need to be tested because it is considered appropriate to be added in concrete.

The polypropylene fiber used is characterized as a monofilament. The properties from this fiber are shown in Table \ref{polypropylene} \cite{garcez2009investigaccao}. Was utilized to the additive superplasticizer \textit{Suplast Rodo 52017}, which increases the concrete fluidity and helps on the reduction of water consumption.

\begin{table}[ht]
\centering
\caption{Polypropylene fiber properties.}
\label{polypropylene}
\begin{tabular}{l c} 
\toprule
\textbf{Property} & \textbf{Value} \\
\midrule
Length (mm) & 10\\
Diameter ($\mu m$) & 12\\
Tensile strength (MPa) & 850\\
Elongation at fracture (\%) & 21\\
Elasticity module (GPa) & 6\\
Alkali resistance & Excellent\\
\bottomrule
\end{tabular}
\end{table}

In the first environment were used two concrete mixes are based on \citeonline{fagundesavaliaccao}, being them: 1:4 (cement: coarse) with 10\% replacement of coarse for sand and water-to-cement ratio (w/c) of 0.35; 1:5 with 10\% replacement of coarse for sand and w/c of 0.35; The third concrete mix was based in \citeonline{monteiro2010concreto}, being it 1:3, without sand and with w/c of 0.32. These concrete mixes are shown in Table \ref{initialtraces}.

\begin{table}[ht]
\centering
\caption{Initial concrete mixes of permeable concrete.}
\label{initialtraces}
\begin{tabular}{l c c c} 
\toprule
\textbf{Material} & \textbf{Concrete mix 1}  & \textbf{Concrete mix 2} & \textbf{Concrete mix 3}\\
\midrule
Cement:gravel & 1:3 & 1:4 & 1:5\\
Sand (\%) & 0 & 10 & 10\\
Water-to-cement ratio & 0.32 & 0.35 & 0.35\\
Additives (ml) & 30 & 40 & 50\\
\bottomrule
\end{tabular}
\end{table}

Of these experiments, concrete mix 2 was chosen as cement mix reference because it got better results in flexure tensile strength, which is the main parameter to be analyzed in this research, as is determined in NBR 16416 \cite{araujo2015nbr}.

In the second environment, were made a concrete mix using the second mix, with the addition of polypropylene fibers, being them:

\begin{itemize}
\item 600 g/m$^3$ fiber addition;

\item 1800 g/m$^3$ fiber addition;

\item 3000 g/m$^3$ fiber addition.

\end{itemize}

Considering all the materials utilized in the process, the permeable concrete was exposed to tests where was analyzed the parameters: permeability, compression resistance, and flexure tensile strength.

As NBR 16416 \cite{araujo2015nbr}, permeable pavements, regardless of their casing type, must have a permeability coefficient bigger than 10$^{-3}$ m/s. This coefficient was evaluated in the laboratory, being tested the casing layer and the layer as a whole.

\begin{figure}[ht]
\vspace{5mm}
  \subfloat{
	\begin{minipage}[c][\width]{
	   0.5\textwidth}
	   \centering
	   \includegraphics[width=0.8\textwidth]{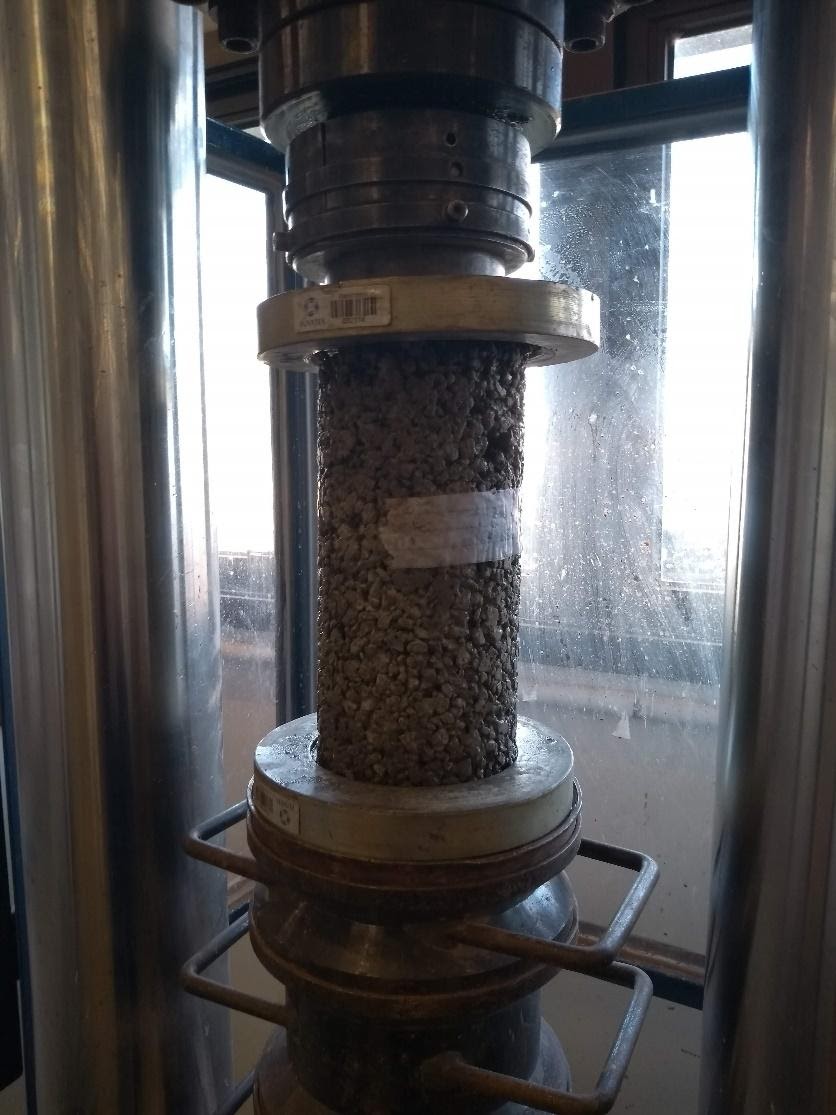}
	   \caption*{Testing the compression resistance.}
	   \label{fig:ensaios_compression_resistance}
	\end{minipage}}
 \hfill 	
  \subfloat{
	\begin{minipage}[c][\width]{
	   0.5\textwidth}
	   \centering
	   \includegraphics[width=0.8\textwidth]{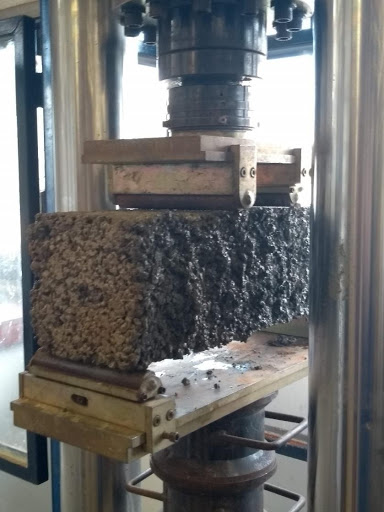}
	   \caption*{Testing the flexure tensile strength.}
	   \label{fig:ensaios_flexure_tensile_strength}
	\end{minipage}}
\vspace{7mm}
\caption{Resistance tests being performed in the specimens.}
\label{fig:ensaios}
\end{figure}

The compression resistance test was made as NBR 15805 \cite{nbr158052015concrete}. In this test were molded cylindrical specimens, with the dimensions of 10 cm $\times$ 20 cm (diameter $\times$ height). Was molded two specimens to each initial concrete mix (a total o six specimens) and four specimens to each concrete mix with the addition of fibers (a total of twelve specimens) that were broken after 28 days. Was realized the increasing density of the specimens by three layers of 15 blows each one. The test is shown in figure \ref{fig:ensaios_compression_resistance}.

The flexure tensile strength was made as NBR 12142 \cite{nbr121422010concrete}. In this test were molded prismatic specimens, with the dimensions of 15 cm $\times$ 15 cm $\times$ 50 cm (height $\times$ width $\times$ depth). In this test, the compaction of the specimens was made by a socket of 2.5 kg, which fell from a height of 30.5 cm on all the surfaces, being effectively 75 blows in each layer and realized in two layers. The test is shown in figure \ref{fig:ensaios_flexure_tensile_strength}. These specimens were tested after 28 days.

%% file: sections/3_results.tex
\section{Results}

Two stages were carried out, where in the first stage, the concrete mixes 1:3, 1:4, and 1:5 of conventional permeable concrete were concreted, while in the second stage, the concrete mix 1:4 were determined as a reference concrete mix, and new concretes were made with the addition of polypropylene fibers. In both stages, permeability, compression resistance, and flexure tensile strength tests were performed. 

\subsection{Permeability}

\begin{figure}[ht]
\centering
\includegraphics[width=\linewidth]{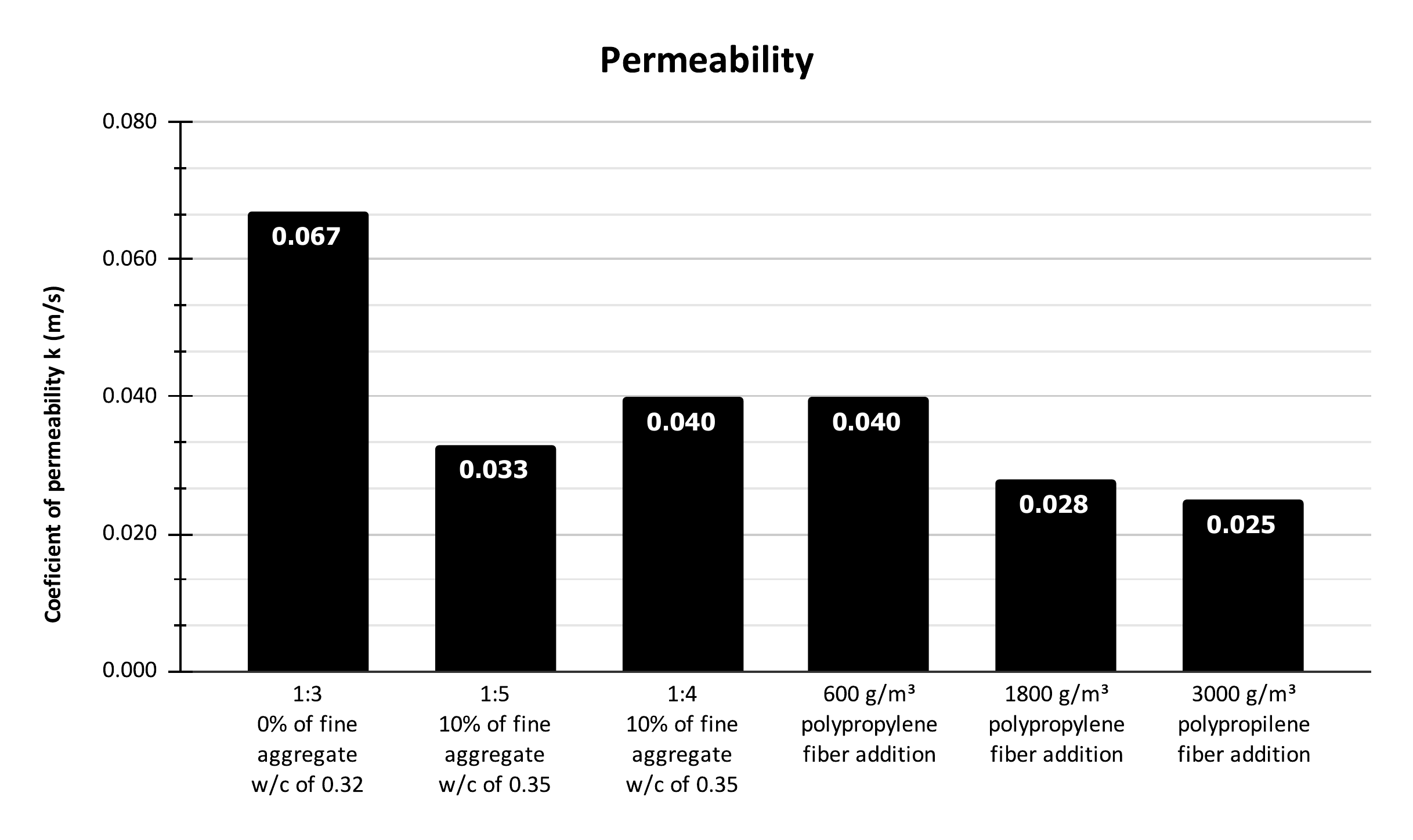}
\caption{Results of conventional permeable concrete and permeable concrete with the addition of polypropylene fiber permeability.}
\label{permeabilidade}
\end{figure}

In the graph in figure \ref{permeabilidade}, it can be verified that all concrete mixes met the permeability requirement conform to NBR 16416 \cite{araujo2015nbr}. In the initial concrete permeability is visualized that the concrete mix 1:3, which did not contain small aggregate in its composition, obtained a higher permeability coefficient of 0.067 m/s, precisely because there are more voids among the concrete materials. The concrete mixes 1:4 and 1:5, which contained fine aggregate in its composition showed values below (0.040 and 0.033, respectively) because there is a greater filling of material with sand, leaving fewer voids in the concrete.

By comparing the 1:4 concrete mix without the addition of fibers and with the addition of polypropylene fibers, it can be seen that the concrete mix using 600 g/m$^3$ kept its permeability coefficient intact, with a value equal to 0.040 m/s. Thus, the use of fiber did not alter the permeability of permeable concrete.

However, the concrete mixes that used 1800 g/m$^3$ and 3000 g/m$^3$ of polypropylene fiber had their permeability coefficients decreased from 0.040 m/s, of the reference concrete mix, to 0.028 m/s and 0.025 m/s, respectively. This change is linked to the use of the fibers because at the time the concreting was done, it can be verified that the fibers made the crushed pieces join each other, becoming agglomerated and reducing the number of voids.

\subsection{Compression Resistance}

\begin{figure}[ht]
\centering
\includegraphics[width=\linewidth]{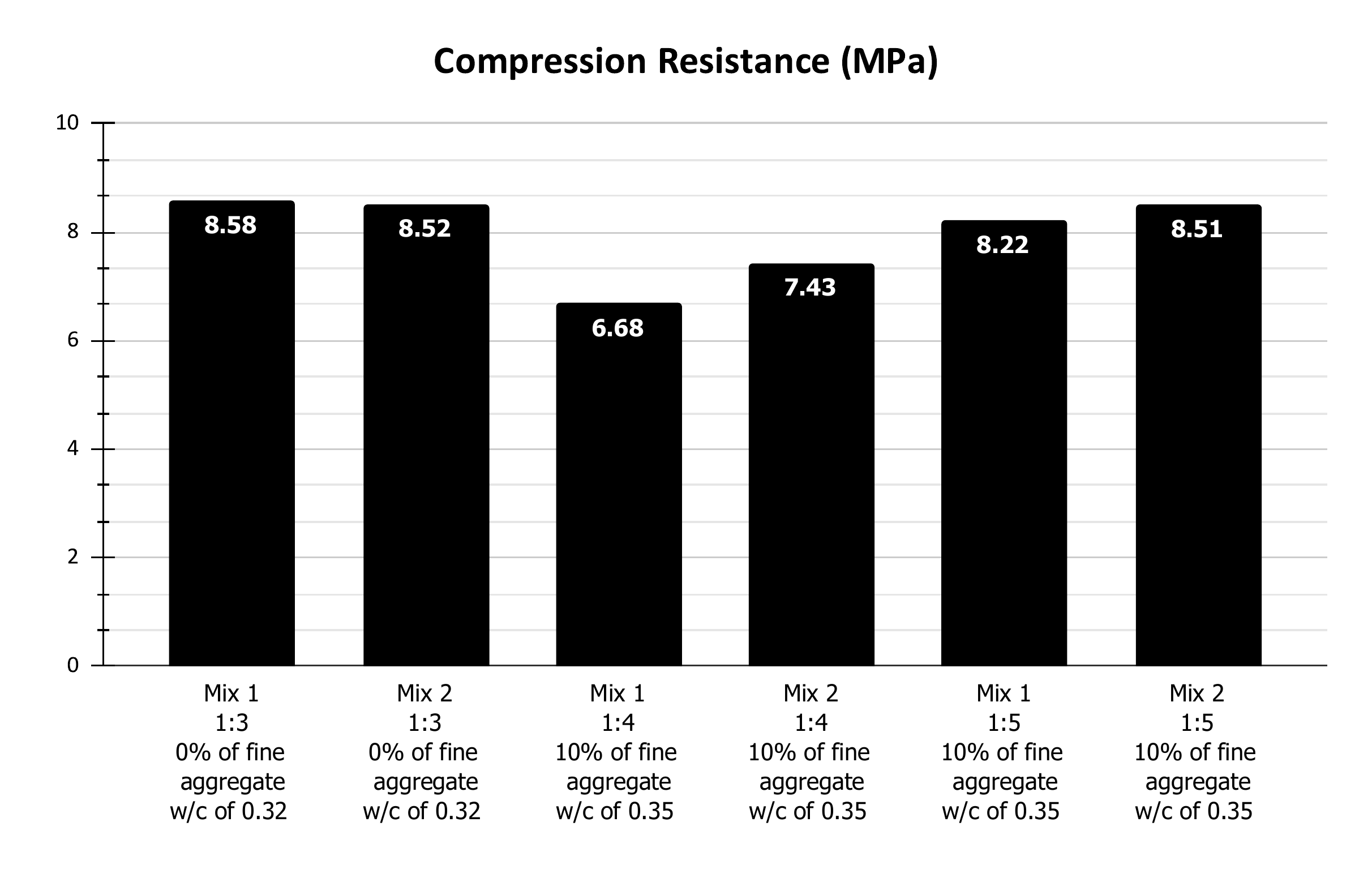}
\caption{Results of the compression resistance of initial concrete mixes.}
\label{compressao1}
\end{figure}

\begin{figure}[ht]
\centering
\includegraphics[width=\linewidth]{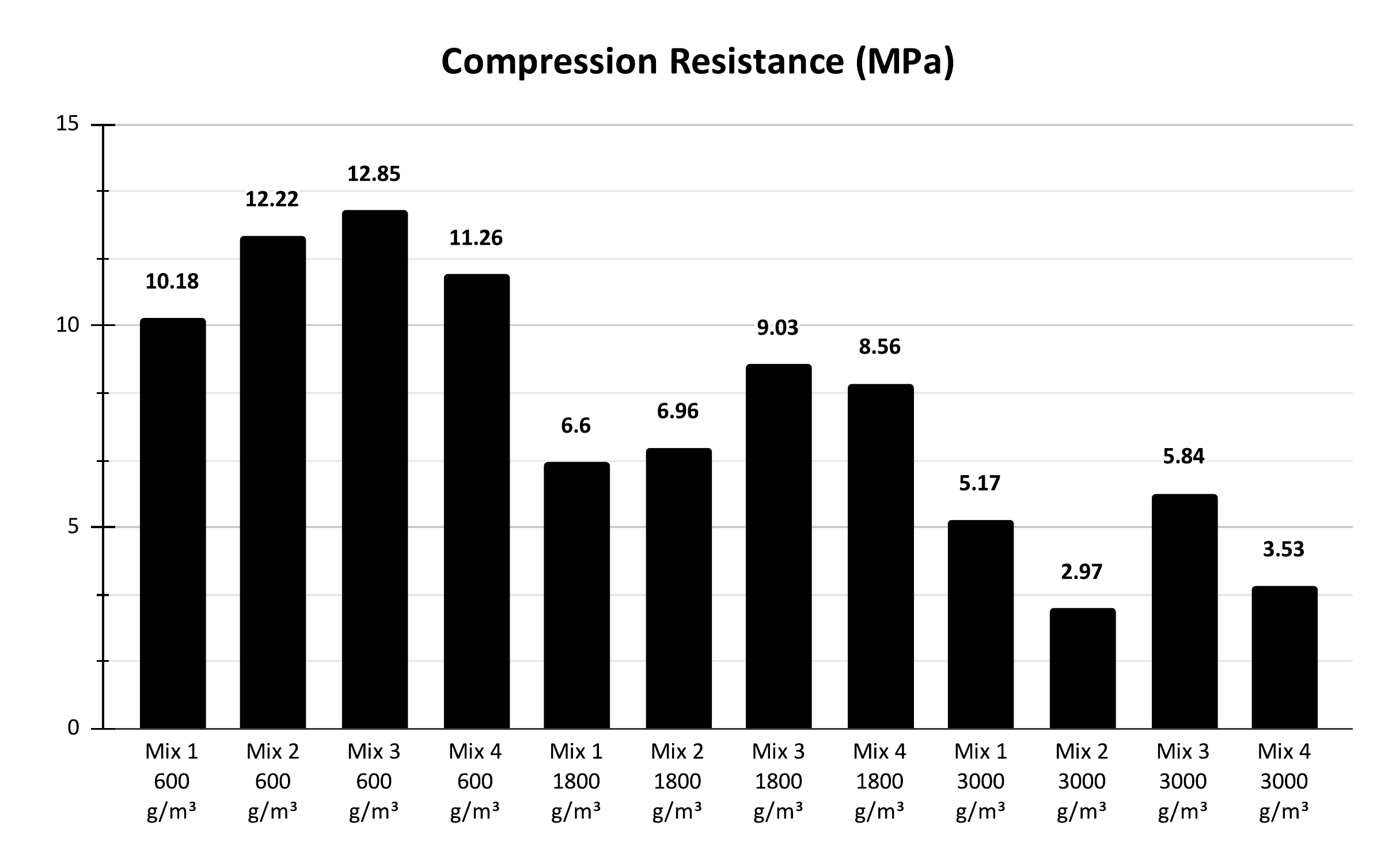}
\caption{Results of compression resistance with the addition of polypropylene fibers.}
\label{compressao2}
\end{figure}

The compression resistance tests of the initial permeable concrete were performed on two specimens for each composition of concrete analyzed. The results are shown in the graph in figure \ref{compressao1}. For the second stage, 4 specimens were developed for each new concrete mix. After the concreting period, the specimens were broken, generating the results shown in the graph in figure \ref{compressao2}. The standard does not establish a minimum value for the compression test with cylindrical specimens with permeable concrete. Therefore, the best analysis to be made is of conventional permeable concrete with added fibers.

The greatest discrepancy of values found for compression resistance in the initial concrete mixes occurred in 1:4, where the values obtained were 6.68 MPa and 7.43 MPa, which may have been caused by the manual thickening of the specimens. In concrete mixes 1:3 and 1:4, the difference in values found for the compression resistance was smaller, being: 8.58 MPa and 8.52 MPa; 8.22 and 8.51 MPa.

Comparing the 1:4 concrete mix without the addition of polypropylene fibers and with the addition of these, it was verified that in the addition of 600 g/m$^3$ of fiber, the concrete obtained an improvement in its results of compression resistance. The original concrete mix obtained 7.43 MPa potential resistance as a result, while with the addition of fibers, 12.85 MPa of potential resistance was found. Thus, it can be verified that this addition improves the performance of the permeable concrete because all the results obtained were greater than the initial concrete mix.

The 1:4 concrete mix with the addition of 1800 g/m$^3$ of polypropylene fiber obtained the potential resistance of 9.03 MPa. As can be observed, this result was above the initial result of 7.43 MPa. Therefore, this addition should be better studied because taking into consideration these results. This concrete mix improves the mechanical property of the concrete concerning its compression resistance.

For the 1:4 concrete mix with the addition of 3000 g/m$^3$ of polypropylene fiber, dispersed results were obtained for the compression resistance being its potential resistance of 5.84 MPa. Besides the results not keeping close values, they were below the original value that was 7.43 MPa. Therefore this addition is not recommended for permeable concrete.

\begin{figure}[ht]
\centering
\includegraphics[width=\linewidth]{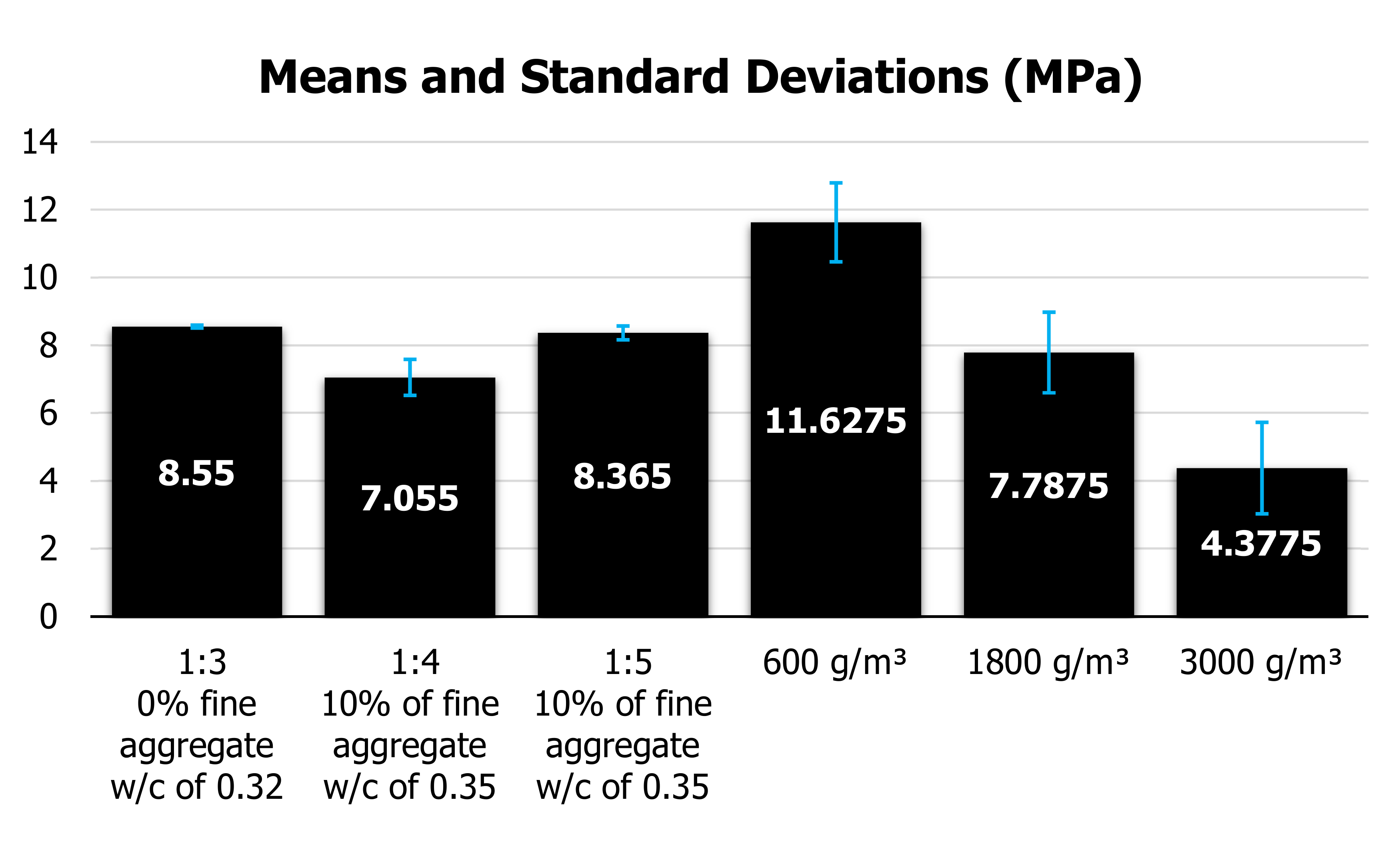}
\caption{Means and standard deviations of compression resistance concrete mixes.}
\label{std1}
\end{figure}

The graph in figure  \ref{std1} shows the calculated means according to the values found in the compression resistance test from the permeable concrete, realized in the two phases of the research, as its respective standard deviation. In this case, it was possible to be observed that the composition of concrete with the addition of polypropylene fibers in the values of 600 g/m$^3$ and 1800 g/m$^3$, had its resistance increased in comparison with the reference composition of concrete of 1:4. On the other hand, the composition of concrete with the addition of 3000 g/m$^3$ of fiber reached lower resistances than its reference composition of concrete.

\subsection{Flexure tensile strength}

\begin{figure}[ht]
\centering
\includegraphics[width=\linewidth]{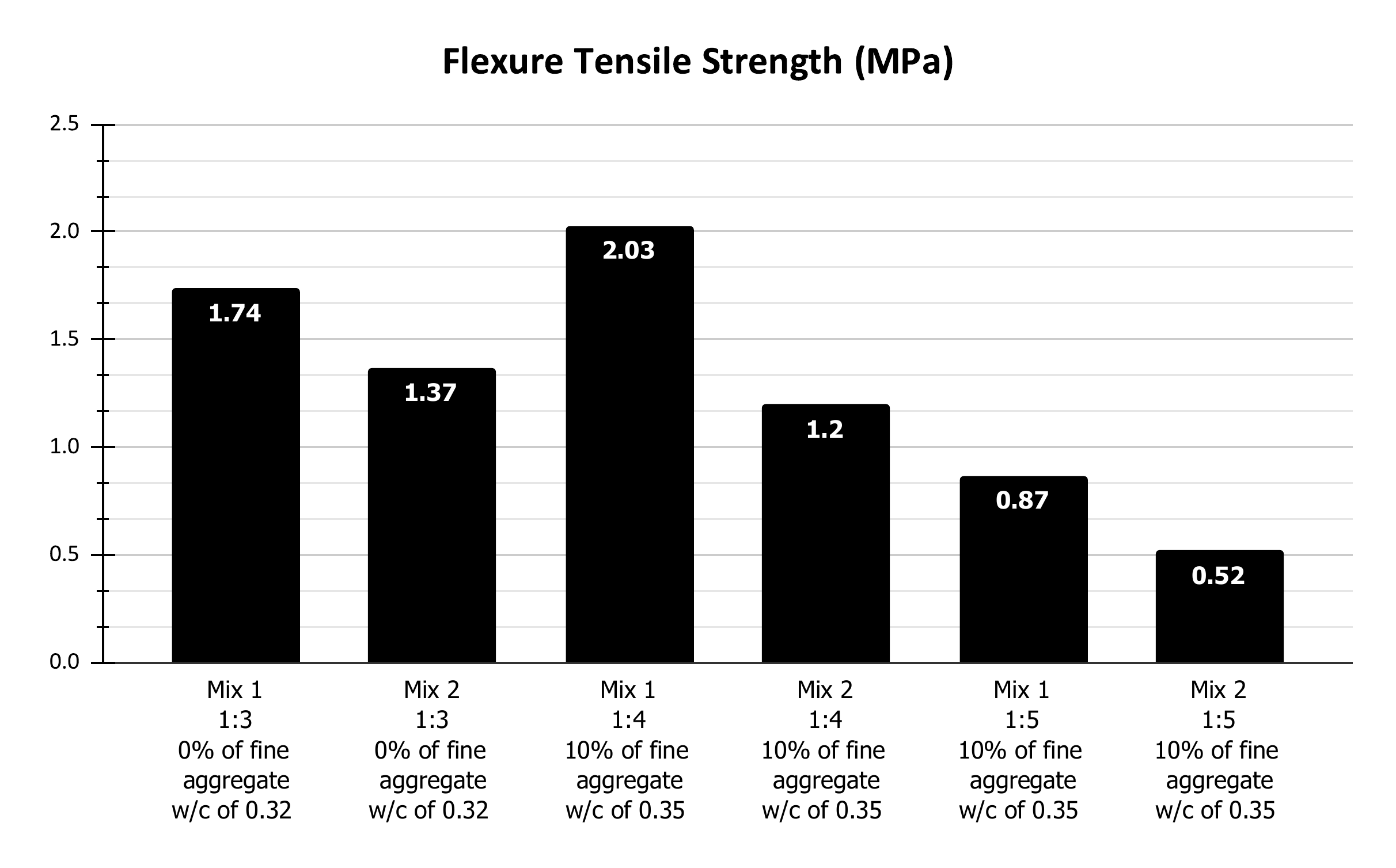}
\caption{Results of flexure tensile strength of initial concrete mixes.}
\label{tracao1}
\end{figure}

Flexure tensile strength tests of the initial permeable concrete was performed with two specimens for each concrete mix analyzed. The results are shown in the graph in figure \ref{tracao1}.

\begin{figure}[ht]
\centering
\includegraphics[width=\linewidth]{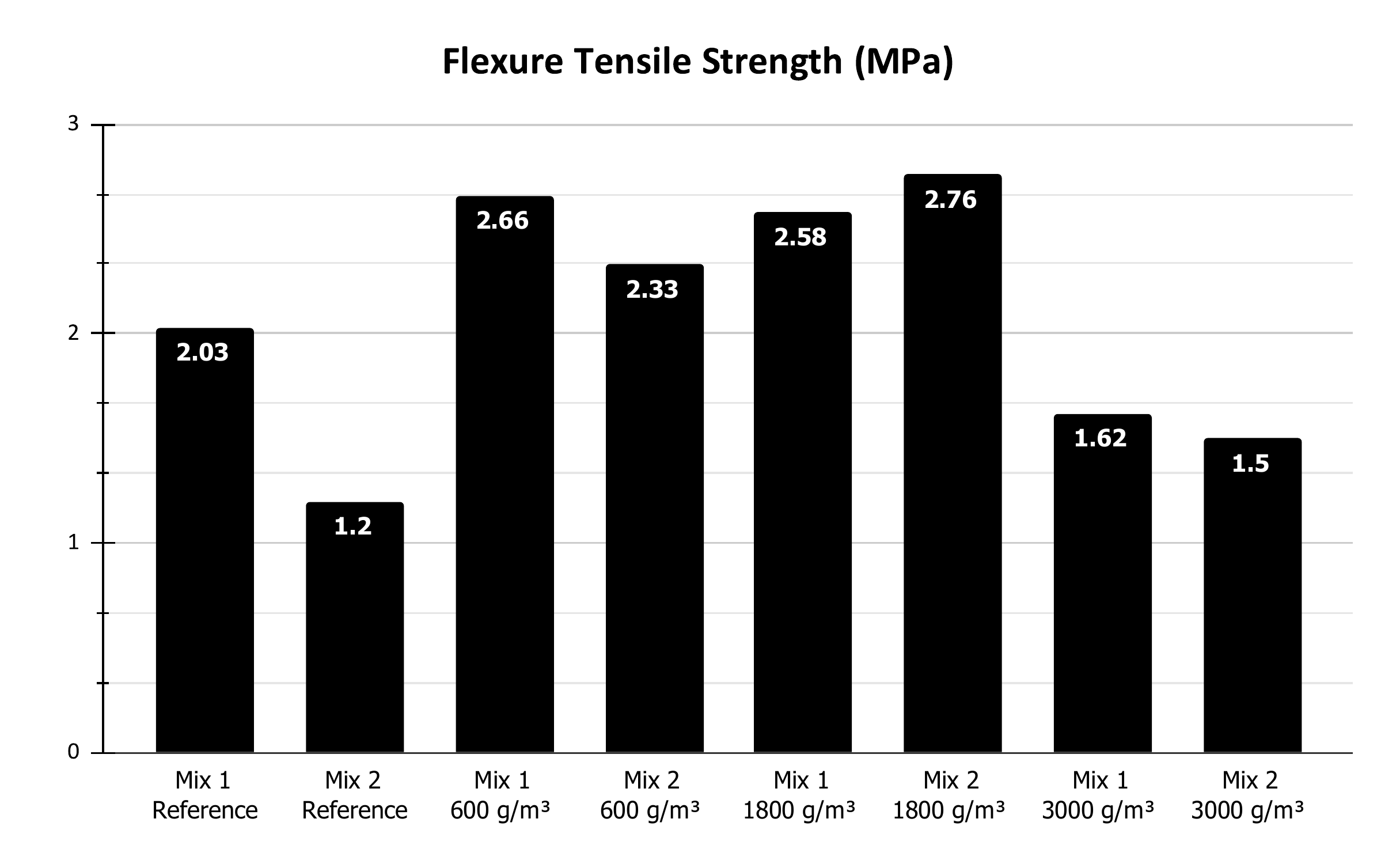}
\caption{Results of flexure tensile strength of the concrete mixes with the addition of polypropylene fibers.}
\label{tracao2}
\end{figure}

After these results were obtained, new concrete was made for the second stage, where the two specimens of each composition of concrete were broken up and the results demonstrated in the graph in figure \ref{tracao2}.

According to NBR 16416 \cite{araujo2015nbr}, for permeable concrete parts, being molded concrete plates, the minimum flexure tensile strength must be 2.0 MPa. In this test, not all concrete mixes have achieved this resistance.

For the initial concrete mixes of permeable concrete, the values found for flexure tensile strength were well dispersed, so that the following results were found: 1.74 MPa and 1.37 MPa (concrete mix 1:3); 2.03 MPa and 1.20 MPa (concrete mix 1:4); 0.87 MPa and 0.52 MPa (concrete mix 1:5). As can be observed only the 1:4 concrete mix presented a value above the minimum established by the standard, which had as potential resistance 2.03 MPa. Thus, this concrete mix was used for the molding of the concrete mixes with the fiber additions, mainly because none of the other two concrete mixes reached the minimum established by the standard. This result may be related to how the test was performed, as well as any imperfections in the samples.

For the 1:4 concrete mix, with the addition of 600 g/m$^3$ of polypropylene fiber, the two samples reached the minimum required by NBR 16416 \cite{araujo2015nbr} and obtained the potential resistance of 2.66 MPa. If compared with the result obtained by the reference concrete mix of potential resistance of 2.03 MPa, the concrete had an improvement in its mechanical performance related to its flexure tensile strength.

In a 1:4 concrete mix, adding 1800 g/m$^3$ of polypropylene fiber, the two samples met NBR 16416 \cite{araujo2015nbr} and obtained a potential resistance result of 2.76 MPa. Compared to its initial concrete mix without the fibers, it obtained an improvement in mechanical performance, besides presenting the best results obtained.

The 1:4 concrete mix, being added 3000 g/m$^3$ of polypropylene fiber, shown in its two samples values below the specified in NBR 16416 \cite{araujo2015nbr}, reaching potential flexure tensile strength of 1.62 MPa. The results obtained were the worst of all concrete mixes with polypropylene fibers, and they were below their initial concrete mix value without the use of 2.03 MPa fibers. It is worth mentioning that, when the specimens of this concrete mix were broken, they did not break in half, demonstrating that the polypropylene fiber held the two parts of the concrete piece, being necessary some hits on the concrete until it broke in two, as seen in figure \ref{CORPOS}.

\begin{figure}[ht]
\centering
\includegraphics[width=\linewidth]{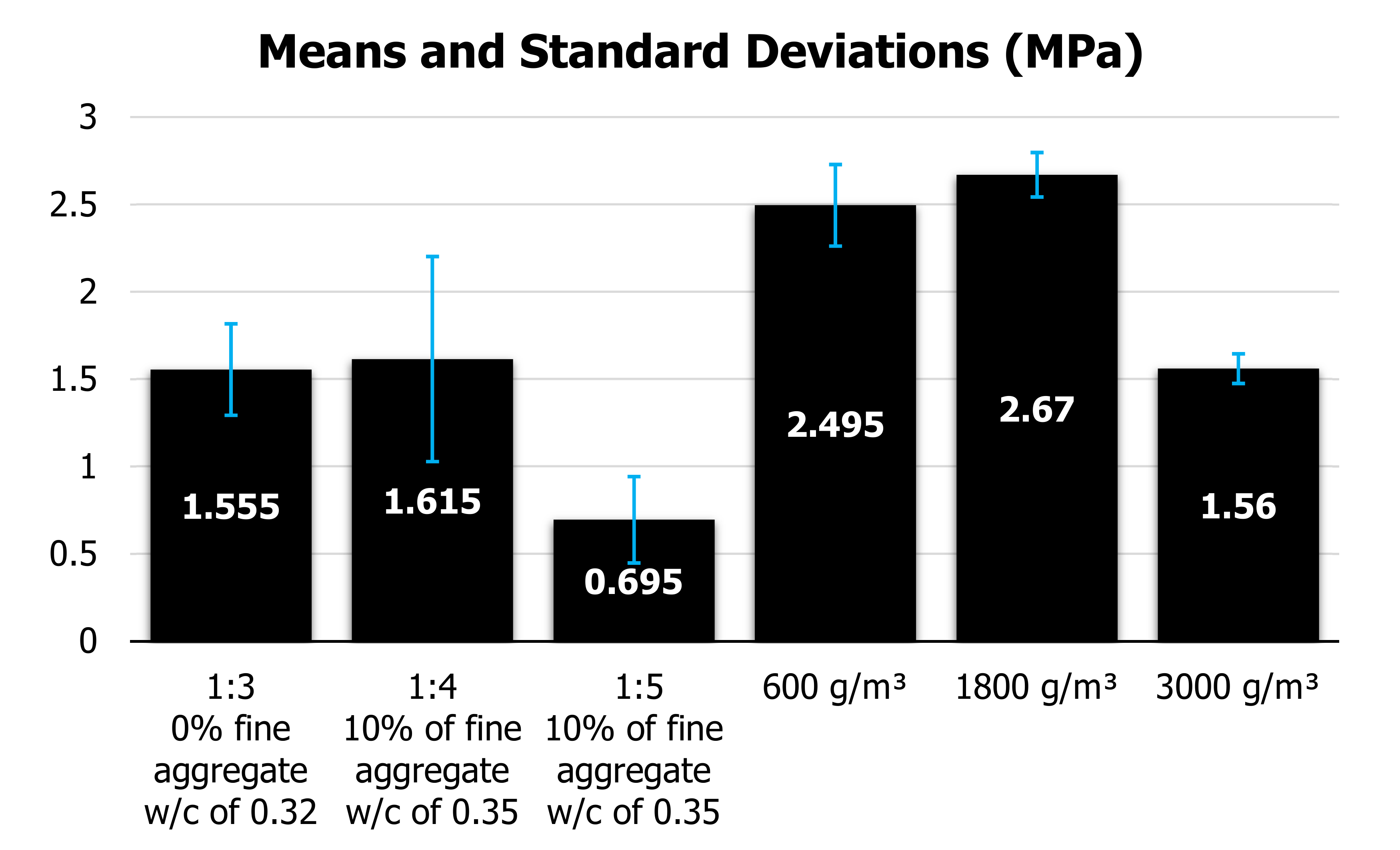}
\caption{Means and standard deviations of flexure tensile strength concrete mixes.}
\label{std2}
\end{figure}

The graph in figure \ref{std2} shows the result of calculated means according to the values found in the flexure tensile strength tests from the permeable concrete, realized in the two phases of the research as its respective standard deviation. According to the graph, it was possible to be observed that the two first compositions of concrete with the addition of polypropylene fibers (600 g/m$^3$ and 1800 g/m$^3$) improved its performances, in this way, increasing its resistances, being the composition of concrete with 1800 g/m$^3$ to achieve the higher resistance. In contrast, the composition of concrete with the addition of 3000 g/m$^3$ had resistance results below the reference composition of concrete of 1:4.

%% file: sections/4_conclusion.tex
\section{Conclusion}

In the hardened state, all the permeable concrete made have obtained a permeability index higher than the minimum established by NBR 16416 \cite{araujo2015nbr} of 10$^{-3}$. It can be verified that the addition of polypropylene fibers in permeable concrete alters the permeability of the concrete mildly. Thus, it can be used, so as not to lose its essence of draining water.

In the compression resistance test, the permeable concrete that reached the highest value was the concrete mix 1:4, with the addition of 600 g/m$^3$ of polypropylene fiber being its value 12.85 MPa. One aspect to consider is that the concrete that was added 3000 g/m$^3$ and reached a resistance of 5.85 MPa, reached a resistance lower than the reference concrete mix which was 7.43 MPa.

In the flexure tensile strength test of permeable concrete, the highest result was the concrete mix 1:4 with the addition of 1800 g/m$^3$, which reached 2.76 MPa and reached the minimum required by the standard. In this test, the addition of polypropylene fiber of 3000 g/m$^3$ that reached a resistance of 1.62 MPa, was below the reference concrete resistance equal to 2.03 MPa and was below the minimum resistance required by the standard.

According to the results obtained during the work, the addition of polypropylene fiber of 600 g/m$^3$ in the permeable concrete increases the resistance of the reference concrete mix concerning the compression test. For the flexure tensile strength test, the addition of 1800 g/m$^3$ of polypropylene fiber increased the results compared to the reference concrete mix.

For the 1:4 concrete mix of permeable concrete with the addition of polypropylene fibers of 3000 g/m$^3$, the results obtained in the compression resistance and the flexure tensile strength were below the reference concrete mix. Therefore it is not advisable to use it to increase its mechanical resistance.

Through the results obtained, it can be observed that the use of polypropylene fibers in permeable concrete must be controlled since the line with the highest addition of fibers obtained the lowest results in its resistances. This can be explained by the adhesion of the fibers to the rest of the permeable concrete aggregates, which contain a greater number of voids, which are occupied by the excessive amount of fibers, not generating resistance to the concrete. This makes the concrete more expensive and less effective for draining water.

On the other hand, the line with the least addition of fibers, despite presenting better results in the tests of compression resistance, its results in flexure tensile strength were below the minimum established in the standard, making this not recommended for being used in permeable concrete.

Thus, among those tested in this research, the best concrete mix for permeable concrete to be used with the addition of polypropylene fiber is the 1:4 with the addition of 1800 g/m$^3$, which uses an intermediate amount of fibers, because it reached the highest flexure tensile strength, and reached the value established by the standard.